\documentclass[12pt]{article}

\usepackage[a4paper,textwidth=17.5cm,textheight=25cm]{geometry}

\usepackage{url}
\usepackage{graphicx}
\usepackage{hyperref}
\usepackage{array}
\usepackage{caption}
\usepackage{multirow}
\usepackage{lscape}
\graphicspath{ {images/} }
\usepackage[round,authoryear]{natbib}
\makeatletter
\renewcommand{\@biblabel}[1]{#1.}
\makeatother
\usepackage{mathtools}
\usepackage{xcolor}
\hypersetup{
    colorlinks,
    linkcolor={black!50!black},
    citecolor={blue!50!black},
    urlcolor={blue!80!black}
}

\usepackage{fixmath}
\usepackage{float}
\usepackage{amsmath,amssymb}
\usepackage{soul}
\usepackage{algcompatible}
\usepackage{algorithm}
\usepackage{algpseudocode}
\usepackage{pdfpages}
\usepackage{setspace}
\usepackage{enumerate}
\usepackage{multirow}
\usepackage{authblk}
\usepackage{titlesec}
\usepackage{accents}
\titleformat{\section}
  {\normalfont\fontsize{12}{12}\bfseries}{\thesection}{0.5em}{}

\titleformat{\subsection}
  {\normalfont\fontsize{12}{12}\bfseries}{\thesubsection}{0.5em}{}

\titleformat{\subsubsection}
  {\normalfont\fontsize{12}{12}\bfseries}{\thesubsubsection}{0.5em}{}

\titleclass{\subsubsubsection}{straight}[\subsection]

\newcounter{subsubsubsection}[subsubsection]
\renewcommand\thesubsubsubsection{\thesubsubsection.\arabic{subsubsubsection}}

\titleformat{\subsubsubsection}
  {\normalfont\fontsize{12}{12}\bfseries}{\thesubsubsubsection}{0.5em}{}
\titlespacing*{\subsubsubsection}
{0pt}{3.25ex plus 1ex minus .2ex}{1.5ex plus .2ex}

\makeatletter
\renewcommand\paragraph{\@startsection{paragraph}{5}{\z@}%
  {3.25ex \@plus1ex \@minus.2ex}%
  {-1em}%
  {\normalfont\normalsize\bfseries}}
\renewcommand\subparagraph{\@startsection{subparagraph}{6}{\parindent}%
  {3.25ex \@plus1ex \@minus .2ex}%
  {-1em}%
  {\normalfont\normalsize\bfseries}}
\def\toclevel@subsubsubsection{4}
\def\toclevel@paragraph{5}
\def\toclevel@paragraph{6}
\def\l@subsubsubsection{\@dottedtocline{4}{7em}{4em}}
\def\l@paragraph{\@dottedtocline{5}{10em}{5em}}
\def\l@subparagraph{\@dottedtocline{6}{14em}{6em}}
\makeatother

\newcounter{algsubstate}
\renewcommand{\thealgsubstate}{\alph{algsubstate}}
\newenvironment{algsubstates}
  {\setcounter{algsubstate}{0}%
   \renewcommand{\State}{%
     \stepcounter{algsubstate}%
     \Statex {\footnotesize\thealgsubstate:}\space}}
  {}
 
\begin{document}
\setlength{\abovedisplayskip}{.4cm}
\setlength{\belowdisplayskip}{.4cm}
\title{\vspace{-.6cm} \large \textbf {Bayesian adaptive N-of-1 trials for estimating population and individual treatment effects}} \vspace{-.8cm}
\author[]{\normalsize Senarathne SGJ\thanks{Corresponding author: jagath.gedara@hdr.qut.edu.au} }
\author[2]{Overstall, AM}
\author[]{McGree JM}
\affil[]{\small School of Mathematical Sciences, Science and Engineering Faculty, Queensland University of Technology, Brisbane, Australia\vspace{0.3cm}}
\affil[2]{\small 
Southampton Statistical Sciences Research Institute, 
University of Southampton,
Southampton, UK}
\date{}
\maketitle \vspace{-.8cm}

\begin{center} {\textbf{ABSTRACT}} \end{center}

This article proposes a novel adaptive design algorithm that can be used to find optimal treatment allocations in N-of-1 clinical trials. This new methodology uses two Laplace approximations to provide a computationally efficient estimate of population and individual random effects within a repeated measures, adaptive design framework. Given the efficiency of this approach, it is also adopted for treatment selection to target the collection of data for the precise estimation of treatment effects. To evaluate this approach, we consider both a simulated and motivating N-of-1 clinical trial from the literature. For each trial, our methods were compared to the multi-armed bandit approach and a randomised N-of-1 trial design in terms of identifying the best treatment for each patient and the information gained about the model parameters. The results show that our new approach selects designs that are highly efficient in achieving each of these objectives. As such, we propose our Laplace-based algorithm as an efficient approach for designing adaptive N-of-1 trials.

 
\section{Introduction}

Research in evidence-based medicine is increasingly moving towards informing individualised clinical care.  This has led to renewed interest in N-of-1 clinical trials as they provide the strongest level of evidence for individual decisions \citep{Guyatt2000}. In N-of-1 trials, patients undergo a series of treatment periods called cycles where each patient receives each treatment (active or placebo) sequentially in a randomised order. This allows each treatment to be trailed on each patient, enabling patients to act as their own control. A major benefit of such a design is that individual and population treatment effects can be estimated through hierarchical modelling approaches as proposed by \cite{Zucker1997}. However, one potential drawback of such a design is that it remains fixed throughout the entire trial. This is potentially limiting as data collected throughout the trial are not used to target informative treatment allocations. For this purpose, we propose a novel adaptive design algorithm for the selection of treatments that target the estimation of both population and individual treatment effects. Such an approach will therefore provide more informative and efficient N-of-1 trials, ensuring the right treatment is selected for the right patient in personalised care.

N-of-1 trials are typically used to test new and/or competing treatments for chronic diseases that are relatively stable over time.  In typical N-of-1 clinical trials, treatments do not permanently change the disease or condition, and therefore, once off treatment, the patient will return to their underlying stable state (after a sufficiently long wash-out period).  Such features may seem restrictive in practice, however, N-of-1 trials have been applied widely to inform clinical care for, for example, arthritis, asthma, insomnia, attention deficit hyperactivity disorder, hypertension, sleep disturbance, and fatigue from cancer \citep{Pope2004, COXETER2003, Nikles2006, Alemayehu2017, NIKLES2019, MITCHELL2015289}. Most notably, N-of-1 trials will typically recruit far fewer patients than randomised controlled trials making them suitable for smaller patient cohorts such as those with rare diseases \citep{Stunnenberg2018,Bellgard2019}. Further, given individual treatment effects can be estimated, N-of-1 trials are suitable when there is significant variability in response to treatment as seen, for example, in response to the treatment of chronic pain \citep{Germini2017}.

To extend the N-of-1 trial design, we consider a Bayesian adaptive design framework as sequential learning through time fits naturally within this framework, and is a framework where uncertainty about, for example, the preferred treatment for a given patient is handled most rigorously. A variety of different adaptive design algorithms have been proposed in the literature (see \cite{ryan2016} for a recent review). Such algorithms often use Markov chain Monte Carlo (MCMC) samplers, importance sampling or sequential Monte Carlo (SMC) methods to update prior information at each iteration of the adaptive design process \citep{Palmer1998,Weir2007, mcgree2012adaptive, McGree2016}. However, such approaches are not appropriate for designing N-of-1 trials as they are either too computationally expensive and/or do not allow both population and individual treatment effects to be estimated.  As such, new statistical methods are needed that allow N-of-1 trials to be efficiently and adaptively designed.

For adaptive design of N-of-1 trials, we consider a Laplace approximation to both the log-likelihood and the posterior distribution of the parameters \citep{Lin1996,SKAUG2006699}. As will be shown, this provides an accurate approximation to the posterior distribution of the population parameters and individual random effects, and can be derived efficiently, without relying on computationally burdensome Monte Carlo methods. The standard Laplace approximation to the posterior distribution has been considered previously \citep{Lewi2009,long2015laplace,Senarathne2020} but has been limited to independent data settings, and, such as, is not appropriate for N-of-1 trials. Thus, we propose a new adaptive design and inference algorithm for N-of-1 trials including an extension to the Laplace approximation for mixed effects models for use in Bayesian design. To select treatments, we aim to maximise the Kullback-Leibler (KL) divergence \citep{kullback1951information} between the prior and the posterior distribution of the population and individual random effect parameters. Such an approach targets the collection of data expected to provide the most information about the parameters. To evaluate this approach, two alternative designs are considered. The first is based on the multi-armed bandit (MAB) design \citep{Thompson1933, Scott2010,Villar2015} where treatments are randomised based on the probability of being the preferred treatment for a given individual.  Of note, our Laplace approximation allows the MAB design to be employed efficiently at the individual level through the availability of estimated individual random effects. Secondly, we compare our designs with those based on a randomised N-of-1 trial design where treatments are randomly selected with equal probability. 

The outline of the paper is as follows. In Section 2, a motivating N-of-1 trial is introduced, along with the statistical model used to analyse the data from this trial. Our Bayesian adaptive design framework is then defined in Section 3. In Section 4, we show how to efficiently approximate the log-likelihood and the posterior distribution of the parameters. To illustrate our methodologies, Section 5 focuses on two examples of aggregated N-of-1 trials. The paper concludes with a discussion of key findings and suggestions for future research.

\section{Motivating Example} \label{Sec:Sec2}

The majority ($60\%-90\%$) of advanced cancer patients experience fatigue, with such fatigue being related to cancer treatment or the disease itself \citep{Vogelzang19974,Lawrence2004}. Cancer-related fatigue (CRF) often persists after the end of treatment and can last for months, or even years \citep{Bower2006}. As reported in previous studies, CRF is more severe and persistent than normal fatigue caused by lack of sleep or overexertion, and has a negative impact on work, social relationships, and daily activities \citep{curt2000, Bower2014}. Despite this, and the fact that such fatigue has significant detrimental effects on the quality of life of cancer patients, in most cases it is under-treated as most patients consider fatigue a symptom to be endured \citep{Vogelzang19974}. 

Methylphenidate (MPH), a psychostimulant, is a commonly prescribed medication for the treatment of CRF. However, studies of this medication have yielded conflicting evidence about it's effectiveness in treating CRF. Little to no effect was observed by \cite{MINTON2011}, \cite{KERR2012} observed significant improvement in advanced prostate cancer patients. Collectively, the results of such studies indicated that the effectiveness of MPH on CRF varied depending on the condition of the patient and cancer type. Such variation motivated the consideration of an N-of-1 clinical trial of MPH by \cite{MITCHELL2015289}. The main goals of the trial were to estimate: (1) the population treatment effect of MPH on CRF in patients with advanced cancer; (2) individual treatment effects; and (3) how variable individual treatment effects are within a population of advanced cancer patients.  In this paper, we consider this study as motivation for our methodological developments to enable adaptive N-of-1 trials to be efficiently designed.

\subsection{Data collection and selection}

\cite{MITCHELL2015289} conducted a series of N-of-1 trials on $43$ patients over three cycles. In each cycle, patients were assigned to both MPH (treatment) and placebo in a randomised order. To ensure patients (and clinicians) were blinded throughout the study, both the treatment and placebo were administered in capsules that were identical in appearance and taste. To measure CRF, the Functional Assessment of Chronic Illness Therapy-Fatigue (FACIT-F) subscale was used. This is a survey comprised of $13$ questions where each response is measured on a five-point Likert scale. A total score (which is the primary outcome) is calculated as the sum across all responses, with higher scores indicating less fatigue.

As reported by \cite{MITCHELL2015289}, 24 patients completed the trial. Among these, 22 patients completed all six treatment periods (i.e.\ yielded complete data from three cycles of two treatments), and these data were considered in this paper to estimate population and individual treatment effects, and the variability of treatment effects within and between patients. 

\subsection{Modelling aggregated N-of-1 trials}

The analysis of aggregated N-of-1 clinical trial data can be undertaken within a mixed effects modelling framework.  Such an approach not only allows population effects to be estimated (akin to that provided by randomised clinical trials) but also allows individual treatment effects to be estimated. \cite{Zucker2010} provide the foundations for this for N-of-1 clinical trials through a generalised linear mixed model (GLMM) specified as follows:

\begin{equation}
  g(\text{E}[y_{ijk}|\boldsymbol{b_{i}},d_{ijk}]) = (\beta_0+b_{0i}) + (\beta_1+b_{1i})d_{ijk},
   \label{Eq:model1}
\end{equation}

where $\text{E}[y_{ijk}|\boldsymbol{b_{i}},d_{ijk}]$ denotes the $j^\text{th}$ expected observation of the $i^\text{th}$ patient at the $k^\text{th}$ cycle, $\boldsymbol{\beta}=(\beta_0,\beta_1)$ are the population parameters, $\boldsymbol{b_{i}}=(b_{0i},b_{1i})$ are the random effects for the $i^\text{th}$ patient, $d_{ijk}$ is the treatment allocation for the $j^\text{th}$ observation of the $i^\text{th}$ patient in the $k^\text{th}$ cycle and $g(.)$ is a link function which maps between the linear predictor and the space of the expected observation. Here, $d_{ijk}$ can take the value 1 or 0, depending on the treatment allocation (treatment=1, placebo=0). In this model, random effects $b_{0},b_{1}$ are assumed to follow a Normal distribution with zero means and variances $\omega_0$ and $\omega_1$, respectively. Further, given the random effects $\boldsymbol{b_{i}}$, $y_{ijk}$ is distributed according to an exponential family distribution as follows:

\[
    f(y_{ijk}|\boldsymbol{b_{i}})=\exp{\Bigg [\frac{y_{ijk}\zeta_{ij}-B(\zeta_{ij})}{A(\phi)} + C(y_{ijk},\phi) \Bigg ]}
\]

\noindent where, $A$, $B$ and $C$ are known functions, $\phi$ is the dispersion parameter of the distribution $f(\cdot|\cdot)$, and $\zeta_{ij}$’s are known as the canonical parameters. When, for example, $y_{ijk}$ has a Normal distribution (conditional on $\boldsymbol{b_{i}}$ and $d_{ijk}$), the variance of $y_{ijk}$ is $\phi = \sigma^2$; the residual variance.

After estimating the parameters in the above model, it can be used to assess population and individual treatment efficacy. Here, $\beta_0$ represents the population level average response when receiving the placebo, and $\beta_1$ represents the population level difference between the average response when receiving the treatment compared to placebo.  For estimating the individual treatment and placebo effects, the relevant individual random effect values are added to the population effects. As such, the average placebo and treatment effects for the $i^\text{th}$ patient can be expressed as $(\beta_0+b_{0i})$ and $(\beta_0+b_{0i}+\beta_1+b_{1i})$, respectively.   

\section{Adaptive design} \label{Sec:Sec3}

Consider an adaptive N-of-1 clinical trial where the goal is to determine with respect to a placebo: (1) the effectiveness of a given treatment in the population; (2) the effectiveness of a given treatment for each individual; and (3) how variable the effectiveness of treatment is in the patient cohort. Without loss of generality, we assume that such an adaptive trial can be constructed by considering treatment selection for each cycle for each patient, iteratively.  That is, initially treatments within the first cycle for the first patient are determined.  Then, once data have been collected from this cycle, treatments for the first cycle for the second patient will be determined, and so on.  Of course, in practice, patients are recruited at different times, thus this ordering may not be exactly how a given trial would proceed.  However, as will be seen, our proposed methodology does not depend on this adaptive structure, and is actually flexible enough to handle the large variety of design problems that may be encountered in real N-of-1 trials. Within our adaptive approach, treatments are allocated without constraint. That is, our approach allows allocating any treatment combination for a given patient within a given cycle. The notion of a cycle is only retained so that comparisons can be made with standard N-of-1 trials. If we suppose there are a total of $N$ number of patients in the study and each patient receives one of the $N$ treatments (including placebo) in each cycle for a total of $K$ cycles.

Then, the design $\boldsymbol{d_{1:NMK}}$ can be expressed as follows:

\[
    \boldsymbol{d_{1:NMK}}= (d_{111},d_{121},\ldots,d_{ijk},\ldots,d_{NMK})^T.
\]

For clarity, consider two treatments (active and placebo) which will be assigned to $N=5$ patients within a single $(K=1)$ cycle. Given there is only a single cycle, the subscript for $K$ can be omitted, and we can define the design $\boldsymbol{d_{1:NM}}$ as follows:

\[
    \boldsymbol{d_{1:NM}}= (d_{11},d_{12},\ldots,d_{15},d_{21},\ldots,d_{25})^T.
\]

As there are no constraints in our treatment allocations, each of the above design points $d_{ij}$ is either active treatment or placebo. This extends naturally to more than a single cycle, comparing more than two treatments and enrolling more than five patients into the trial.

Within an adaptive design framework, the Bayesian inference problem is to evaluate the joint posterior distribution of the population parameters $\boldsymbol{\theta}=(\boldsymbol{\beta},\sigma^2,\omega_1,\omega_2)$ (defined here, for example, for a Normally distributed response) and the random effects $\boldsymbol{b}$ at each iteration of the adaptive design. This posterior is typically analytically intractable, therefore we are required to sample from or approximate the following:

\[
    p(\boldsymbol{\theta,b|y_{1:ijk},d_{1:ijk}}) \propto p(\boldsymbol{y_{1:ijk}|\beta},\sigma^2,\boldsymbol{b,d_{1:ijk}})p(\boldsymbol{b|\omega})p(\boldsymbol{\theta}),
\]

where $\boldsymbol{\omega}=(\omega_1,\omega_2)$ denotes the random effect variances, and $p(\boldsymbol{y_{1:ijk}|\beta},\sigma^2,\boldsymbol{b,d_{1:ijk}})$ denotes the conditional likelihood of observing data $\boldsymbol{y_{1:ijk}}=(y_{111},\ldots,y_{ijk})^T$ from design $\boldsymbol{d_{1:ijk}}$ given the population parameters $\boldsymbol{\theta}$ and the random effects $\boldsymbol{b}$, for $i=1,\ldots,N$, $j=1,\ldots,M$ and $k=1,\ldots,K$.  Further, $p(\boldsymbol{b|\omega})$ is the distribution of the random effects $\boldsymbol{b}$ conditional on the parameters $\boldsymbol{\omega}$, and $p(\boldsymbol{\theta})$ is the prior distribution of the population parameters.

The Bayesian adaptive design problem can then be stated as selecting $d_{(ijk)+1}$ at each iteration. Here, depending on the iteration, the design $d_{(ijk)+1}$ can be either $d_{i\{j+1\}k}$, $d_{\{i+1\}jk}$ or $d_{ij\{k+1\}}$. That is, in each iteration of this design algorithm, we select a treatment for either the same patient, the next patient or for the first patient in the study to start the next treatment cycle. For selecting which treatment to administer, a utility function is defined to reflect the aim of the study which we assume, based on the three goals stated at the start of Section 3, is parameter estimation. In general, we denote the utility function as $U(d,z,\boldsymbol{\theta,b} |\boldsymbol{d_{1:ijk},y_{1:ijk}})$, where $z$ is the response obtained from running design $d$.  However, as $z$, $\boldsymbol{\theta}$ and $\boldsymbol{b}$ are unknown, the expectation is taken with respect to the joint distribution of these random variables based on the posterior distribution from the previous iteration. This yields the following expected utility:

\begin{equation}
\begin{split}
U(d|\boldsymbol{d_{1:ijk},y_{1:ijk}}) & =E_{z,\boldsymbol{\theta,b}|\boldsymbol{d_{1:ijk},y_{1:ijk}}}[U(d,z,\boldsymbol{\theta,b}|\boldsymbol{d_{1:ijk},y_{1:ijk}})] \\
& = \int\limits_{\boldsymbol{Z}}\int\limits_{\boldsymbol{\theta}}\int\limits_{\boldsymbol{b}} U(d,z,\boldsymbol{\theta,b}|\boldsymbol{d_{1:ijk},y_{1:ijk}}) \\
& \hspace{2 cm}  \times p(z|d,\boldsymbol{\beta},\sigma^2,\boldsymbol{b})p(\boldsymbol{\theta,b|d_{1:ijk},y_{1:ijk}}) \mbox{d}{\boldsymbol{b}} \mbox{d}{\boldsymbol{\theta}} \mbox{d}{z},
\end{split}
\label{eq:ch3_ute} 
\end{equation}

\noindent where the above expected utility $U(d|\boldsymbol{d_{1:ijk},y_{1:ijk}})$ is defined based on a continuous response variable. Extensions to other types of responses are straightforward.

When the utility function does not depend on the population parameters $\boldsymbol{\theta}$ and the random effects $\boldsymbol{b}$, Equation (\ref{eq:ch3_ute}) can be simplified to yield,

\begin{equation}
U(d|\boldsymbol{d_{1:ijk},y_{1:ijk}}) = \int\limits_{\boldsymbol{Z}} U(d,z|\boldsymbol{d_{1:ijk},y_{1:ijk}})p(z|d) \mbox{d}{z},
\label{eq:ch3_ute2} 
\end{equation}

\noindent where $p(z|d)$ is the model evidence.

Thus, at each iteration of the adaptive design process, one seeks to find $d^*_{(ijk)+1} = \text{arg} \; \underset{d \in \mathcal{D}}{\text{max}}~ U(d|\boldsymbol{d_{1:ijk},y_{1:ijk}})$, and this is termed the optimal design.  Unfortunately, the above expression for the expected utility generally cannot be solved analytically, and thus needs to be approximated.  The most common approach for this is Monte Carlo integration through the simulation of prior predictive data as follows:

\begin{equation}
 U(d|\boldsymbol{d_{1:ijk},y_{1:ijk}}) \approx \frac{1}{Q} \sum_{q=1}^Q U(d,z_q\boldsymbol{|d_{1:ijk},y_{1:ijk}}), 
 \label{Eq:Ch3_Monte_ute}
\end{equation}

\noindent where $z_q\sim p(z|d)$.

The adaptive design process described above is outlined in Algorithm \ref{Alg:Design_gen} where initially the prior information about the parameters is defined.  Typically, in N-of-1 trials, this prior information will be uninformative or vague but this need not be the case within our framework. Then, throughout the iterative process, the next optimal design point is found by maximising the expected utility (line 5), and the next data point is collected (line 6) based on this selected optimal design. The prior information about the population parameters and the random effects is then updated based on the information gained from the new data point (line 7). For the examples considered in this desktop study, data cannot actually be collected.  In place of this, we assume data are generated from an underlying model with specified parameter values.  For the motivating study, this underlying model is based on the results from analysing data from \cite{MITCHELL2015289}.

\begin{algorithm}[H]
\small
 \caption{Bayesian adaptive design algorithm for N-of-1 trials}
 	\begin{algorithmic}[1]
 	\State Initialise the prior information $p(\mathbold{\theta,b})$ for the population parameters
        \For{$k=1$ to $K$} 	
            \For{$i=1$ to $N$} 
                \For{$j=1$ to $M$} 
                   \State Find the optimal design point $d_{(ijk)+1}$ by maximising the utility ${U}(d| \boldsymbol{y_{1:ijk},d_{1:ijk}})$
                    \State Collect data point $y_{(ijk)+1}$ at design point $d_{(ijk)+1}$ 
                    
                    \State Update the joint posterior distribution $p(\boldsymbol{\theta,b|y_{1:(ijk)+1},d_{1:(ijk)+1}})$
                \EndFor
            \EndFor
 	    \EndFor
\end{algorithmic}
\label{Alg:Design_gen}
\end{algorithm}  

In considering the adaptive design process as outlined in Algorithm \ref{Alg:Design_gen}, there are two main challenges (at least computationally).  The first is efficiently updating prior information as new data arrive (line 7). Employing methods like MCMC would require re-running this algorithm at each iteration of the adaptive design. Given the high-dimensional nature of the posterior distribution (i.e. many random effects to estimate), this will quickly become computationally infeasible.  Thus, alternative approaches are needed. The second difficulty is evaluating the expected utility function which requires sampling from or approximating a large number of posterior distributions, see Equation (\ref{Eq:Ch3_Monte_ute}). Thus, efficient approaches are needed, and this motivates the development of the new methods proposed in this paper.

\section{Efficient approximation to the joint posterior distribution} \label{Sec:Sec4}

A Laplace approximation is used to efficiently approximate the posterior distribution of the parameters. This approximation is formed by finding the posterior mode and evaluating the inverse of the negative Hessian matrix of the log posterior density at this mode. These two terms form the mean and variance-covariance matrix of a multivariate normal distribution approximation to the posterior distribution.  To describe how this Laplace approximation is formed in this paper, we first show how the posterior mode is located.  For this purpose, we first define the likelihood for random effect models such as those defined in Equation (\ref{Eq:model1}).  For such models, the likelihood function is defined by integrating out the random effects as follows:

\begin{equation}
\begin{split}
   L(\boldsymbol{\theta;y_{1:ijk}}) & =\int\limits_{\boldsymbol{b}} p(\boldsymbol{y_{1:ijk}|\theta,b,d_{1:ijk}})p(\boldsymbol{b|\omega})\text{d}\boldsymbol{b} \\
    & =\int\limits_{\boldsymbol{b}}\exp\{h(\boldsymbol{b,\theta;y_{1:ijk},d_{1:ijk}}) \}\text{d}\boldsymbol{b},
\label{Eq:likelihood}
\end{split}
\end{equation}

\noindent where $h(\boldsymbol{b,\theta;y_{1:ijk},d_{1:ijk}})=\log p(\boldsymbol{y_{1:ijk}|\theta,b,d_{1:ijk}})+ \log p(\boldsymbol{b|\omega})$ is the joint log-likelihood function for the population parameters and the random effects.

For some models, the above integral can be solved analytically.  However, in general, there is no closed-form solution, so an approach to handle the general case is presented. Accordingly, when an analytic solution is not available, an approximation is required. Following the work of \cite{Breslow1993}, a Laplace approximation to the log-likelihood function is

\begin{equation}
\hat{l}(\boldsymbol{\theta;y_{1:ijk}})=-\frac{1}{2}\log|\boldsymbol{H(\boldsymbol{b^*_{\theta}})}|+h(\boldsymbol{b^*_{\theta},\theta;y_{1:ijk},d_{1:ijk}}),
\label{Eq:Laplace1}
\end{equation}

\noindent where 
\begin{equation}
\boldsymbol{b^*_{\theta}}=\underset{\boldsymbol{b}}{\operatorname{arg\,max}}{\; h(\boldsymbol{b,\theta;y_{1:ijk},d_{1:ijk}})} \; \text{and} 
\label{Eq:b_star}
\end{equation}

\begin{equation}    
\boldsymbol{H(\boldsymbol{b^*_{\theta}})} = \frac{\partial^2{\big\{ h(\boldsymbol{b,\theta;y_{1:ijk},d_{1:ijk}}) \big\}}} {\partial\boldsymbol{b}\partial\boldsymbol{b}^{'}}\Big|_{\boldsymbol{b=b^*_{\theta}}}
\label{Eq:Hess_new}
\end{equation}

\noindent is the Hessian matrix evaluated at $\boldsymbol{b^*_{\theta}}$. 

Based on the above approximation to the log-likelihood function, the posterior mode of the population parameters can be found as follows:

\begin{equation}
\boldsymbol{\theta^*} = \underset{\boldsymbol{\theta}}{\operatorname{arg\,max}} \; \hat{l}(\boldsymbol{\theta;y_{1:ijk}})+\log p(\boldsymbol{\theta}).
\label{Eq:log_like_theta}
\end{equation}

\noindent However, from the above formulation, it can be seen that $\boldsymbol{b^*_{\theta}}$ is conditional on $\boldsymbol{\theta}$ and $\boldsymbol{\theta^*}$ is conditional on $\boldsymbol{b^*_{\theta}}$. Hence, to find the posterior mode for both $\boldsymbol{\theta}$ and $\boldsymbol{b}$, conditional optimisation is used. That is, for an initial value of the population parameters, values of the random effects which maximise $h(\boldsymbol{b,\theta;y_{1:ijk},d_{1:ijk}})$ are found. These random effects are then used to approximate the log-likelihood which is subsequently used to find the posterior mode for $\boldsymbol{\theta}$. This process then continues until convergence, upon which $\boldsymbol{\theta^*}$ and $\boldsymbol{b^*}$ denote the mode of the posterior distribution.  It is this mode that is taken as the mean of the multivariate normal approximation to the posterior distribution.

Once the posterior mode has been found, the Hessian matrix at this point can be evaluated and used to form the variance-covariance matrix of the multivariate normal distribution.  For this, we note that the model specified in Equation (\ref{Eq:model1}) assumes the population parameters are independent of the random effects, hence this variance-covariance matrix will be block diagonal. The block corresponding to the random effects can be found via the Hessian matrix in Equation (\ref{Eq:Hess_new}), and the block corresponding to the population parameters can be found as follows:

\begin{equation}           
        \boldsymbol{A(\theta^*)} = \frac{\partial^2\{\hat{l}(\boldsymbol{\theta;y_{1:ijk}})+\log p(\boldsymbol{\theta}) \}}{\partial{\boldsymbol{\theta}}\partial{\boldsymbol{\theta}}^{'}}\Big|_{\boldsymbol{\theta}=\boldsymbol{\theta^*}}. 
        \label{Eq:Hess}
\end{equation}

\begin{algorithm}[H]
\small
 \caption{Laplace approximation for the posterior distribution of population and individual parameters}
 	\begin{algorithmic}[1]
     \State Find $\boldsymbol{\theta}^*$ by maximising Equation (\ref{Eq:log_like_theta}) 
     
     \begin{algsubstates}
        \State For each proposed $\boldsymbol{\theta}$, find $\boldsymbol{b^*_{\theta}}=\underset{\boldsymbol{b}}{\operatorname{arg\,max}} \; h(\boldsymbol{b,\theta;y_{1:ijk},d_{1:ijk}})$ and the Hessian matrix $\boldsymbol{H(b^*_{\theta})}$ (Equations (\ref{Eq:b_star}) and (\ref{Eq:Hess_new}))
        
        \State Update approximation to log-likelihood $\hat{l}(\boldsymbol{\theta;y_{1:ijk}})$, see Equation (\ref{Eq:Laplace1})
     \end{algsubstates}
     
     \State Given $\boldsymbol{\theta}^*$, find  $\boldsymbol{b^*_{\theta^*}}$ and the Hessian matrix $\boldsymbol{H(b^*_{\theta^*})}$ 
  
     \State Approximate $p(\boldsymbol{\theta,b|y_{1:ijk},d_{1:ijk}}) \sim \text{MVN}\big((\boldsymbol{\theta^*,b^*_{\theta^*}}),\boldsymbol{\Omega}\big)$ where $\boldsymbol{\Omega}=\begin{bmatrix} 
        \boldsymbol{-A(\theta^*)^{-1}} & \textbf{0} \\
        \textbf{0} & \boldsymbol{-H(b^*_{\theta^*})^{-1}} 
        \end{bmatrix}$
\end{algorithmic}
\label{Alg:Laplace_post}
\end{algorithm}  

The process for approximating the posterior distribution of the population parameters and the random effects is outlined in Algorithm \ref{Alg:Laplace_post}. To initialise the algorithm, a value for the population parameters is randomly drawn from the prior distribution. Given these values $\boldsymbol{\theta}$, the mode of the random effects and the Hessian matrix at this mode are found (line 1a). These two quantities are used to form a Laplace approximation to the log-likelihood function given in Equation (\ref{Eq:Laplace1}). Next, this approximate log-likelihood function is used to approximate a density that is proportional to the log-posterior distribution of the population parameters. This density is then used to locate the posterior mode $\boldsymbol{\theta}^*$ of the population parameters. Given $\boldsymbol{\theta}^*$, the mode of the random effects and the Hessian matrix at this mode are found (line 2). Finally, the (joint) posterior distribution of the population and individual parameters is approximated using a multivariate normal distribution (line 3). It is worth noting that when the likelihood function has an analytic solution, computation of $\boldsymbol{\theta}^*$ is more straightforward as locating the posterior mode of the population parameters can be undertaken directly without continually updating the approximation to the log-likelihood. 

\section{Simulation studies} \label{Simulation}

Here, two aggregated N-of-1 trials are considered to demonstrate the adaptive design approach proposed in Section \ref{Sec:Sec3} with the approximations given in Section \ref{Sec:Sec4}. Since the main objective of these trials is to determine the preferred treatment at the population and individual patient level, the KL divergence utility \citep{kullback1951information} was implemented for treatment selection. KL divergence is a measure of how different one probability distribution is from another. \cite{Lindy1956} proposed that such a measure should be used in design selection if one is interested in maximising the information gain on model parameters. Thus, we implemented this utility as maximising information will lead to minimising uncertainty about parameter values including estimates of random effects for each patient. This should therefore reduce the uncertainty about which treatment is preferred for each patient. The KL divergence utility is

\begin{equation}
   U(d,z\boldsymbol{|y_{1:ijk},d_{1:ijk}})= \int\limits_{\boldsymbol{\theta}}\int\limits_{\boldsymbol{b}}{p(\boldsymbol{\theta,b|y_{1:ijk},d_{1:ijk}},z,d)}\log\Bigg(\frac{p(\boldsymbol{\theta,b|y_{1:ijk},d_{1:ijk}},z,d)}{p(\boldsymbol{\theta,b|y_{1:ijk},d_{1:ijk}})}\Bigg) \text{d}\boldsymbol{b}\text{d}\boldsymbol{\theta},
    \label{Eq:Ute_est1}
\end{equation} 

\noindent where $p(\boldsymbol{\theta,b|y_{1:ijk},d_{1:ijk}})$ and $p(\boldsymbol{\theta,b|y_{1:ijk},d_{1:ijk}},z,d)$ denote the prior and the posterior distribution of the population parameters and the random effects, respectively.

When both prior $p(\boldsymbol{\theta,b|y_{1:ijk},d_{1:ijk}})$ and the posterior distribution $p(\boldsymbol{\theta,b|y_{1:ijk},d_{1:ijk}},z,d)$ follow a multivariate Normal distributions with mean vectors $(\boldsymbol{\mu_0,\mu_1})$ and covariance matrices $(\boldsymbol{\Sigma_0,\Sigma_1})$, respectively, then the KLD utility can be approximated as follows: 

\begin{equation}
    \hat{U}(d,z\boldsymbol{|y_{1:ijk},d_{1:ijk}})= \frac{1}{2}\Bigg( tr\big(\boldsymbol{\Sigma_0^{-1}\Sigma_1}\big) +(\boldsymbol{\mu_1-\mu_0})^T\boldsymbol{\Sigma_0^{-1}}(\boldsymbol{\mu_1-\mu_0})-\eta+\log\Big(\frac{\text{det}\boldsymbol{\Sigma_0}}{\text{det}\boldsymbol{\Sigma_1}} \Big)\Bigg),
    \label{Eq:Ute_est2}
\end{equation}

\noindent where $\eta$ is the dimension of the two multivariate Normal distributions.

Throughout the examples, one active treatment will be compared to placebo, and hence $M=2$. Given this, treatment will be coded as an indicator variable with `1' denoting active treatment and `0' denoting placebo.  For selecting the optimal design, as there are only two possibilities for design selection, the expected utility for each will be evaluated, and the choice which yields the expected utility value is selected. For comparison, we benchmarked this optimal design approach against the MAB design and a randomised N-of-1 trial design. 

A MAB design refers to a sequential experiment in which the goal is to determine the choice or `arm' that yields the largest `reward'. In the context of a clinical trial, each arm represents a particular treatment (active or placebo), and reward refers to benefit from treatment. The MAB design determines treatment choice by randomising treatment selection based on probabilities which reflect current knowledge about the preferred treatment allocation for each patient in the N-of-1 trial. The adaptive design approach proposed in this paper provides a framework to select MAB designs based on individual specific probabilities, and this offers more flexibility than previous approaches based on population parameters only \citep{Scott2010}. Here, benefit from treatment is based on the posterior predictive distribution for each patient. For example, if a higher response indicates a higher level of disease severity, then reward can be quantified via the following probability:  

\begin{equation}
    p_i(d)=p\big(\mu^i_d=\min(\mu^i_0,\mu^i_1)\big), ~~ \text{for}~ d\in\{0,1\},
\label{Eq:MAB}    
\end{equation} 

\noindent where $\mu^i_d=\text{E}[y_i|d]$ denotes the posterior mean for treatment $d$. Within a Bayesian framework, the reward probability given in Equation (\ref{Eq:MAB}) can be expressed as an expectation of an indicator function. Let $\mathcal{I}_d(\boldsymbol{\theta,b_i})=1$ if $\mu^i_d(\boldsymbol{\theta,b_i})=\min(\mu^i_0(\boldsymbol{\theta,b_i}),\mu^i_1(\boldsymbol{\theta,b_i}))$, and $\mathcal{I}_d(\boldsymbol{\theta,b_i})=0$ otherwise. Then,

\[
     p_i(d)= \int\limits_{\boldsymbol{\theta}}\int\limits_{\boldsymbol{b_i}} \mathcal{I}_d(\boldsymbol{\theta,b_i})p(\boldsymbol{\theta,b_i|y_{1:ijk},d_{1:ijk}}) \text{d}\boldsymbol{b_i} \text{d}\boldsymbol{\theta}.
\]

\noindent Here, the reward probability $p_i(d)$ is evaluated based on the posterior distribution of the population parameters and the random effects of the $i^\text{th}$ patient, obtained from the data $\boldsymbol{y_{1:ijk}}$ collected from all the patients given the treatments $\boldsymbol{d_{1:ijk}}$ up to cycle $k$.

This reward probability can be approximated by drawing a large number of samples from the joint posterior distribution of the population parameters and the random effects of the $i^\text{th}$ patient, and evaluating the following:

\begin{equation}
     \hat{p}_i(d) \approx \frac{1}{Q}\sum_{q=1}^Q{\mathcal{I}_d(\boldsymbol{\theta_q,b_{iq}})}, 
     \label{Eq:reward_prob}  
\end{equation}

\noindent where $(\boldsymbol{\theta_q,b_{iq}}) \sim p(\boldsymbol{\theta,b_i|y_{1:ijk},d_{1:ijk}})$.

For the randomised N-of-1 design, the entire treatment sequence for each patient for all treatment cycles (Eg: $\{1,0,0,1,0,1\}$) can be obtained before starting the experiment. This treatment sequence is generated in such a way that the patient receives both treatment and placebo within each treatment cycle in a random order. As such, even though we use an adaptive design approach, randomised N-of-1 designs do not use the information collected throughout the trial to inform treatment selection.  This is exactly how typical N-of-1 trials are run.

In each of the two examples considered in this paper, a simulation study was undertaken where experiments were sequentially simulated within our design framework. That is, in each cycle, the next data point for the $i^\text{th}$ patient was generated based on the selected optimal design and the assumed model (Equation (\ref{Eq:model1})) with true parameter values (line 6 in Algorithm 1). Once the data have been generated, they will be used to update the prior information of the population and individual parameters. Designs for each patient within a given cycle will be found sequentially such that $M$ data points will be collected for a given patient, then $M$ data points will be collected for the next patient.  Once all patients have completed the current cycle, a new cycle for the initial patient will begin.  This will continue until all patients complete three cycles. For both examples, the prior distribution for the parameters was assumed to be vague, and follow Normal distributions as given in Table \ref{tab:priors}. These prior distributions were selected as they are relatively uninformative on the scale of each parameter, and might typically be implemented in practice for inference. 

\begin{table}[H]
\caption {Prior distributions of the population parameters} \label{tab:priors} \vspace{-.6cm}
\begin{center}
\begin{tabular}{ l l} \hline
 Parameter & Prior distribution\\[0.5ex]  \hline 
 $\beta_0$ & $N\big(0,100^2\big)$\\ [0.4ex] 
 $\beta_1$ & $N\big(0,100^2\big)$  \\ [0.4ex]  
 $\log(\sigma)$ & $N\big(2.5,1.6^2\big)$  \\ [0.4ex]
 $\log( \sqrt{\omega_0})$ & $N\big(2.5,1.6^2\big)$\\ [0.4ex] 
 $\log( \sqrt{\omega_1})$ & $N\big(2.5,1.6^2\big)$ \\ [0.4ex] 
 \hline 
\end{tabular}
\end{center}
\end{table} 

As the results are subject to variability through the simulated data, all simulated studies were repeated 20 times to explore the range of outcomes that could be observed. After the designs were obtained and the corresponding data for each example was generated, the posterior distributions were re-evaluated using a standard MCMC approach. This re-evaluation step was undertaken to remove any potential bias from the Laplace approximation to the posterior distribution. We note, however, that the posterior distributions from MCMC and the Laplace approximation were very similar, see Figure S10 in the online supplementary material showing a comparison of these posterior distributions for the simulated example. In addition, in the supplementary material, we have provided results from these simulation studies where an alternative distribution (i.e. the Poisson distribution) is considered for the response, see Section B. All simulations were carried out using R 3.5.2, and R code to reproduce the results in this paper is available via the following GitHub repository, \href{https://github.com/SenarathneSGJ/Adaptive_N-of-1_trials_design}{https://github.com/SenarathneSGJ/Adaptive{\textunderscore}N-of-1{\textunderscore}trials{\textunderscore}design}.

\subsection{Simulated example}

In this example, we investigate the effect of an active treatment over placebo by running an adaptive N-of-1 trial with 20 patients. Here, the response variable is assumed to follow a Normal distribution with higher response values indicating a higher level of disease severity (e.g. pain as measured on a VAS scale). For each patient, six observations were collected over three treatment cycles, and the above defined three approaches for treatment selection were considered. 

Within this simulation study, four design scenarios were considered, each differing in terms of the parameter values assumed in the underlying generative model. Firstly, we considered a group of patients in which the between-subject variability was much smaller than the within-subject variability of the outcome, and where there was a significant difference between treatment and placebo at the population level. This scenario was considered as a baseline setting, where the remaining scenarios were defined by changing a parameter value in this baseline. Specifically, in Scenario 2, we increased the between-subject variability such that it equalled the within-subject variability of the outcome. Scenarios 3 and 4 were constructed by changing the population treatment effect such that there was a large and no difference (respectively) between treatment and placebo at the population level. These four scenarios are summarised in Table \ref{tab:True_parameters} by defining the parameter values used in each. 

\begin{table}[ht]
\caption {The population parameter values for each design scenario} \label{tab:True_parameters} \vspace{-.6cm}
\begin{center}
\begin{tabular}{c c c c c} \hline
 Parameter & Scenario 1 & Scenario 2 & Scenario 3 & Scenario 4\\[0.5ex]  \hline 
 $\beta_0$ & 25 & 25 & 25 & 25 \\ [0.4ex] 
 $\beta_1$ & -1 & -1 & -3 & 0  \\ [0.4ex]  
 $\sigma^2$ & 9 & 9 & 9 & 9  \\ [0.4ex]
 $\omega_0$ & 2.25 & 9 & 2.25 & 2.25 \\ [0.4ex] 
 $\omega_1$ & 2.25 & 9 & 2.25 & 2.25 \\ [0.4ex] 
 \hline 
\end{tabular}
\end{center}
\end{table} 

For each design scenario, separate simulation studies were conducted. In each simulation study, a set of random effects were generated based on the assumed parameter values, and these were considered as the true effects for each patient in the study, and thus used to generate `real' data throughout each N-of-1 trial. Furthermore, the same set of values were also used to calculate the true treatment effect for each patient, and hence to determine the best treatment (active or placebo) for each patient via the model described in Equation (\ref{Eq:model1}).

\textbf{Results:} Figures  \ref{fig:Log_det_E1}, \ref{fig:Best_Treat_probsE1} and  \ref{fig:Best_Treat_receivedE1} summarise the results from Scenario 1. Here, we first assessed the posterior precision of the population and random effect parameters when data were collected based on the optimal (KLD), MAB and randomised N-of-1 designs. For this comparison, we evaluated the log-determinant of the variance-covariance matrix of the joint posterior distribution after each cycle of the experiment. Figure \ref{fig:Log_det_E1} shows boxplots of the distribution of the log-determinant values of each intermediate posterior distribution (for each cycle) for all simulations. As can be seen, the posterior distributions obtained from optimal designs have lower log-determinant values compared to those obtained from MAB and randomised N-of-1 designs. This indicates that the posterior distributions obtained from the optimal design have higher precision compared to those obtained from either MAB or randomised N-of-1 design. As more data are collected, the MAB design also performs relatively well for estimation when compared to the randomised N-of-1 design.

Next, we assessed the designs in terms of identifying the best treatment allocation (active or placebo) for each patient in the study. For this purpose, we first calculated the true treatment effects based on the true parameter values as explained in Section 2.2. Then, based on the true treatment effects, the best treatment assignment $d_{\text{best}}$ for each patient was identified. Next, we evaluated the probability of identifying the best treatment for each patient based on each of the three designs. For this, we obtained a large number of samples from the joint posterior distribution of the population parameters and the random effects. Then, for each posterior sample, we calculated the individual mean response based on each treatment allocation. The required probability for each patient was then estimated via Equation (\ref{Eq:reward_prob}) with $d=d_{\text{best}}$. These probabilities were evaluated for all independent simulations, and the results were averaged to compare the performance of optimal, MAB and randomised N-of-1 designs.

Figure \ref{fig:Best_Treat_probsE1} shows the probability (with $95\%$ credible intervals) of identifying the best treatment allocation for each patient after each treatment cycle when the optimal, MAB and randomised N-of-1 designs were considered for Scenario 1. According to Figure \ref{fig:Best_Treat_probsE1}, the optimal design has higher probability values (with less uncertainty) compared to the MAB and randomised N-of-1 designs. For patients who preferred placebo over the treatment (negative treatment difference), the median probability values were close to $0.5$ (not 1) for all designs in this scenario. The reason these probabilities are not closer to one is that, under the true model (and therefore what will most likely be inferred from the data), it is likely that for a given patient, the active treatment will be preferred.  Hence, stronger evidence is needed to shift individual effects towards placebo when compared to patients who prefer the active treatment.  This is particularly noticeable in this simulation study as the between subject variability is small compared to the within subject variability meaning patients who favour placebo are much less likely to occur than those who favour treatment.  This can be seen by comparing these results to those from Scenario 2 where the between subject variability is large.  As can be seen, probabilities for patients with similar treatment effect differences are generally closer to 1.  Of note, this is not a feature of our approach to treatment selection, but rather a feature of all designs considered here including the randomised N-of-1 trial design.

Figure \ref{fig:Best_Treat_receivedE1} compares the proportion of times the best treatment was administered for each patient in each cycle when the optimal, MAB and randomised N-of-1 designs were considered. Here, MAB design chose the best treatment for each patient a larger number of times than the other two design approaches. The optimal design also chose the best treatment for each patient a reasonable number of times except for the eleventh patient in the study. As the eleventh patient had the highest true treatment effect difference, the optimal design selected the placebo a large number of times. It is quite reasonable to observe such a difference as the optimal designs focus on learning about parameter values while MAB explores with a preference for the preferred treatment (based on currently available information). For this patient, it was more beneficial (in terms of learning about parameter values) to administer the placebo more often than not.

Similar to design Scenario 1, we compared the performance of the optimal, MAB and randomised N-of-1 designs under the remaining three design scenarios. For all of these three scenarios, the optimal designs were able to precisely estimate the joint posterior distribution of the population parameters and the random effects when compared to MAB and randomised N-of-1 designs (see Figures S1, S2 and S3 in the online
supplementary material). As shown in Figures S4 and S6, when there was large uncertainty about the random effects (Scenario 2) or a small difference between the two treatments at the population level (Scenario 4), it was difficult to determine the best treatment assignment for patients where the difference between treatment and placebo was small. However, when there was a clear difference between the two treatments at the population level (Scenario 3), it was relatively straightforward to determine the best treatment allocation for each patient in the study (Figure S5). In all these scenarios, the optimal design performed better than MAB and randomised N-of-1 designs for determining the best treatment assignment for each patient. Figures S7, S8 and S9 in the online supplementary material compare the proportion of times the best treatment was given to each patient when treatments were assigned using the optimal, MAB and randomised N-of-1 designs under design Scenario 2, 3 and 4, respectively. As can be seen, for all scenarios, MAB design chose the best treatment for each patient more often than the other two design methods. However, this did not translate into providing more information about which treatment is better for each patient (as the optimal design performed best for this).  Such results are expected based on how treatments are selected within each approach.

\begin{center}
\includegraphics[width=14cm]{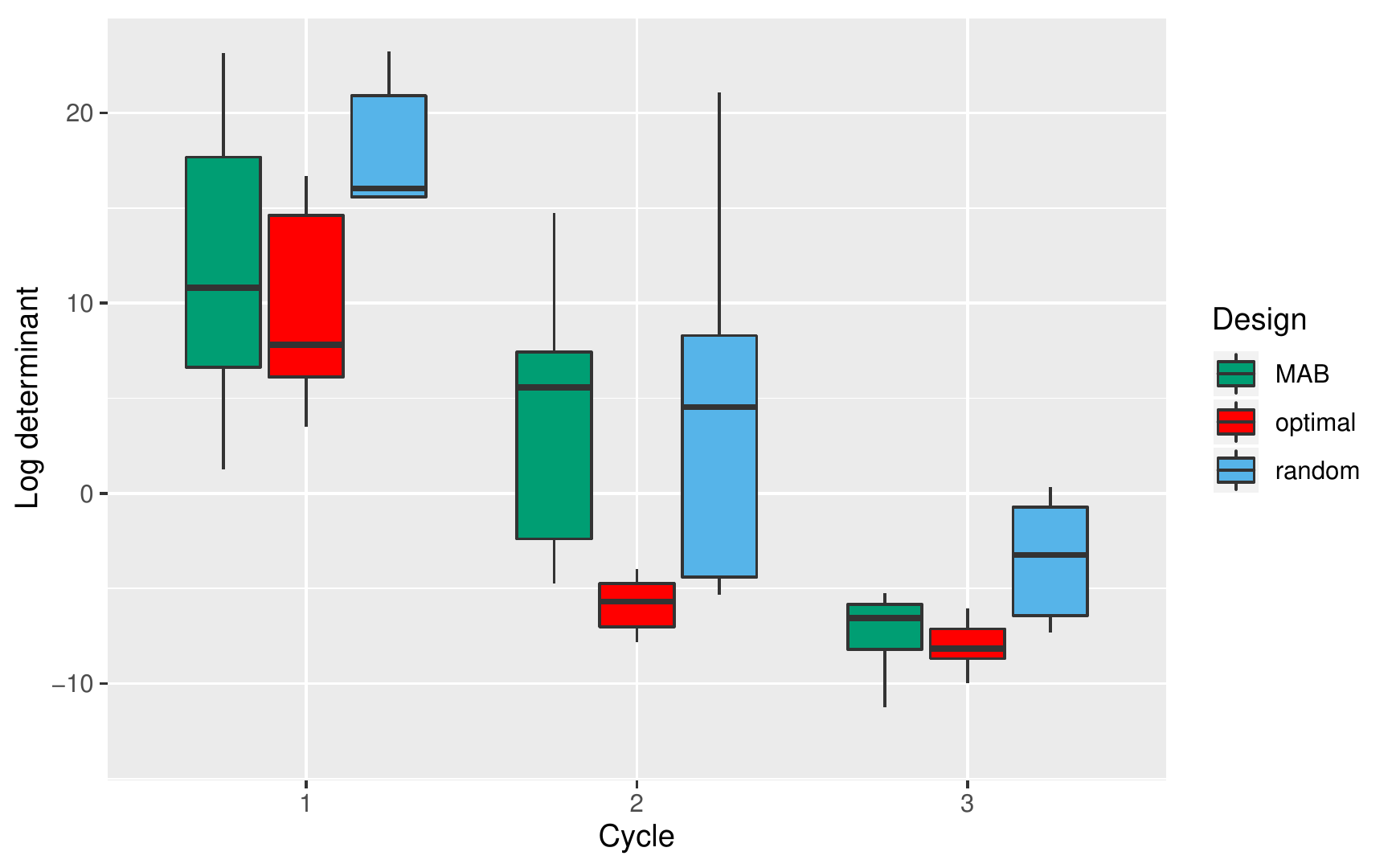} \vspace{-.2cm}
\captionsetup{type=figure}
\captionof{figure}{The boxplot of the distribution of the log-determinant of the posterior variance-covariance matrix for each design after each treatment cycle over 20 simulations from Scenario 1 in Example 1.}
\label{fig:Log_det_E1}
\end{center}

\begin{center}
\includegraphics[width=16cm]{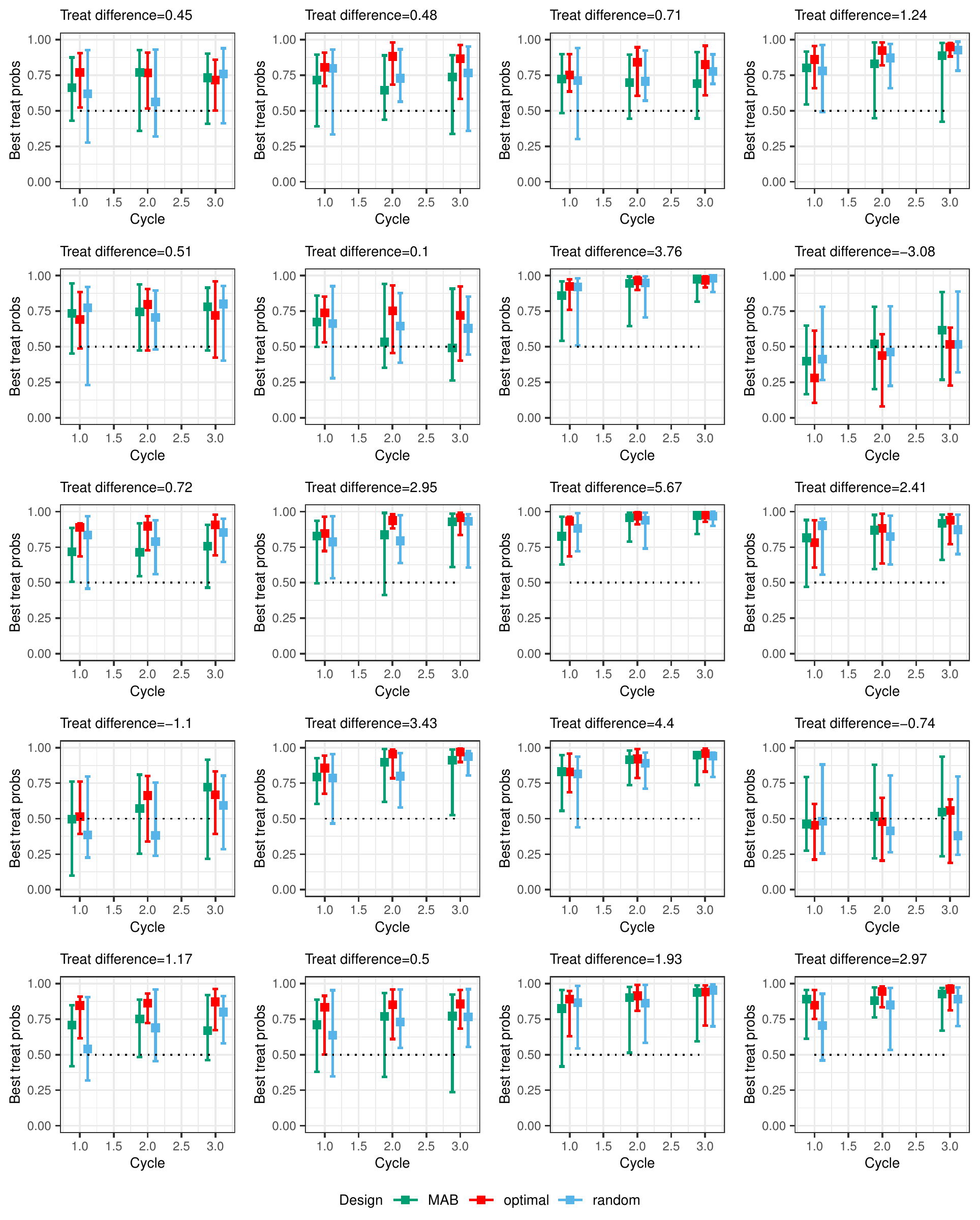} \vspace{-.2cm}
\captionsetup{type=figure}
\captionof{figure}{The probability (with $95\%$ credible intervals) of identifying the best treatment for each patient after each cycle when treatments were assigned using optimal, MAB and randomised N-of-1 designs under Scenario 1 in Example 1.}
\label{fig:Best_Treat_probsE1}
\end{center}

\begin{center}
\includegraphics[width=16cm]{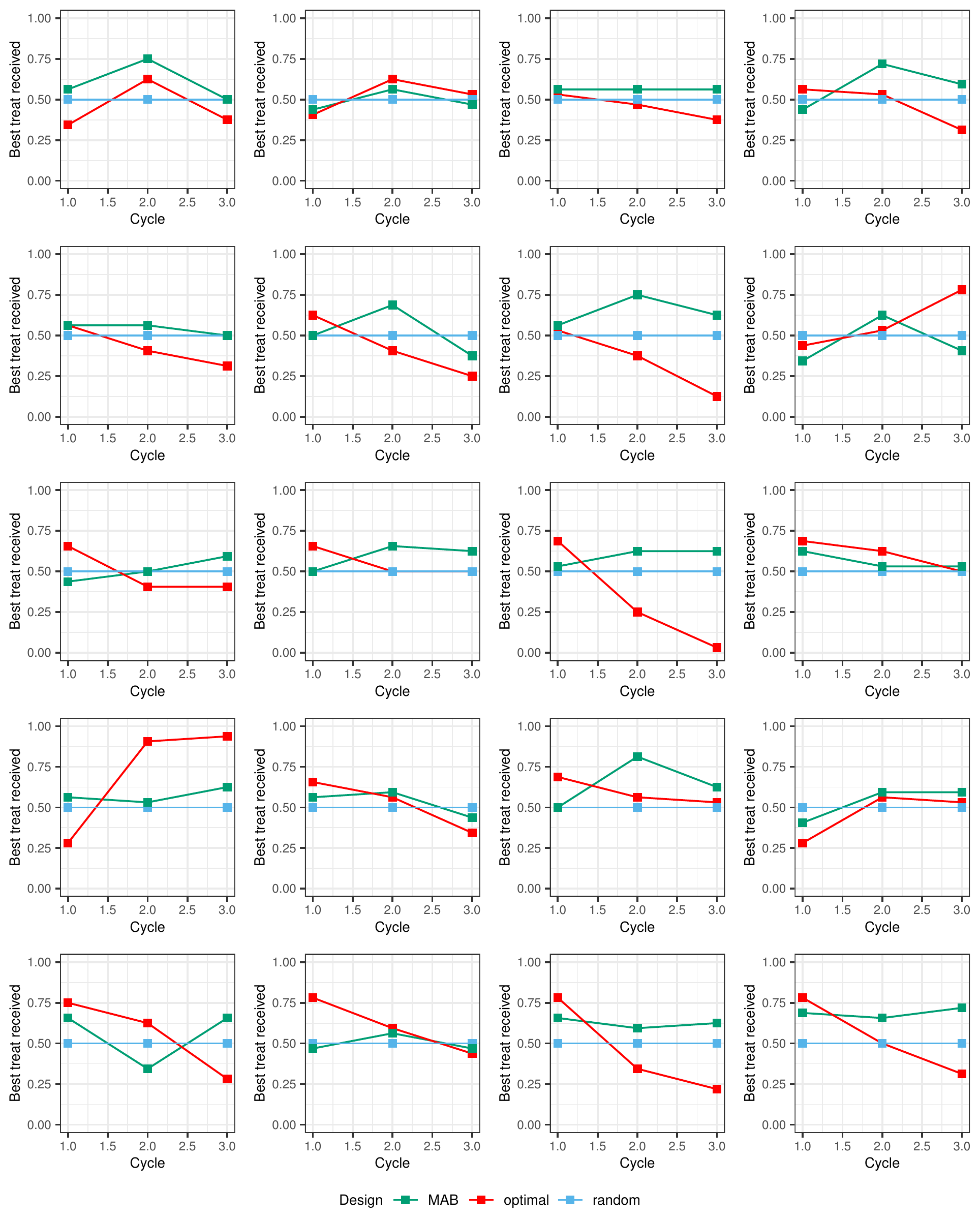} \vspace{-.2cm}
\captionsetup{type=figure}
\captionof{figure}{The proportion of time that their best treatment was received for each patient in each treatment cycle when treatments were assigned using optimal, MAB and randomised N-of-1 designs over 20 simulations for three cycles under Scenario 1 in Example 1.}
\label{fig:Best_Treat_receivedE1}
\end{center}

\subsection{Motivating example}

We now return to our motivating example of MPH trial introduced in Section 2. Based on the data collected from all 22 patients who completed all three cycles in \cite{MITCHELL2015289}, population and individual parameter estimates were found by fitting the model described in Equation (\ref{Eq:model1}). A simulation study was then undertaken with treatment selection based on the optimal, MAB and randomised N-of-1 designs. This simulation study was run exactly as outlined for Example 1, including the prior distributions used (see Table \ref{tab:priors}). Here, the response variable was assumed to follow a log-normal distribution with higher values indicating less fatigue (higher recovery).

\textbf{Results:} For the purpose of comparing the parameter estimation results, log-determinant values of the posterior variance-covariance matrix of the population parameters and the random effects were evaluated and plotted. As shown in Figure \ref{fig:Log_det_E2}, the joint posterior distribution based on the optimal designs have smaller log-determinant values compared to those based on MAB and randomised N-of-1 designs. That is, the optimal design was relatively efficient for parameter estimation. Again, as more design points were collected, MAB design performed relatively well for estimation when compared to the randomised N-of-1 design.

In Figure \ref{fig:Best_Treat_probsE2}, we compare the probability (with $95\%$ credible intervals) of identifying the best treatment assignment for each patient after each treatment cycle when the optimal, MAB and randomised N-of-1 designs were considered for data collection. Similar to the first example, these probabilities were approximated using Equation (\ref{Eq:reward_prob}), but with a different indicator function. Here, the indicator function equals to 1 if the mean response value of the best treatment assignment for a given patient is higher than that of the remaining treatment assignment, and 0 otherwise. As can be seen, the optimal design performed relatively well for this experimental goal when compared to the other two design methods. 

The proportion of times the best treatment was selected for each patient in each treatment cycle is shown in Figure \ref{fig:Best_Treat_receivedE2}. Again, MAB design selected the best treatment for each patient a larger number of times when compared to the other two design methods. \vspace{-.3 cm}

\begin{center}
\includegraphics[width=14cm]{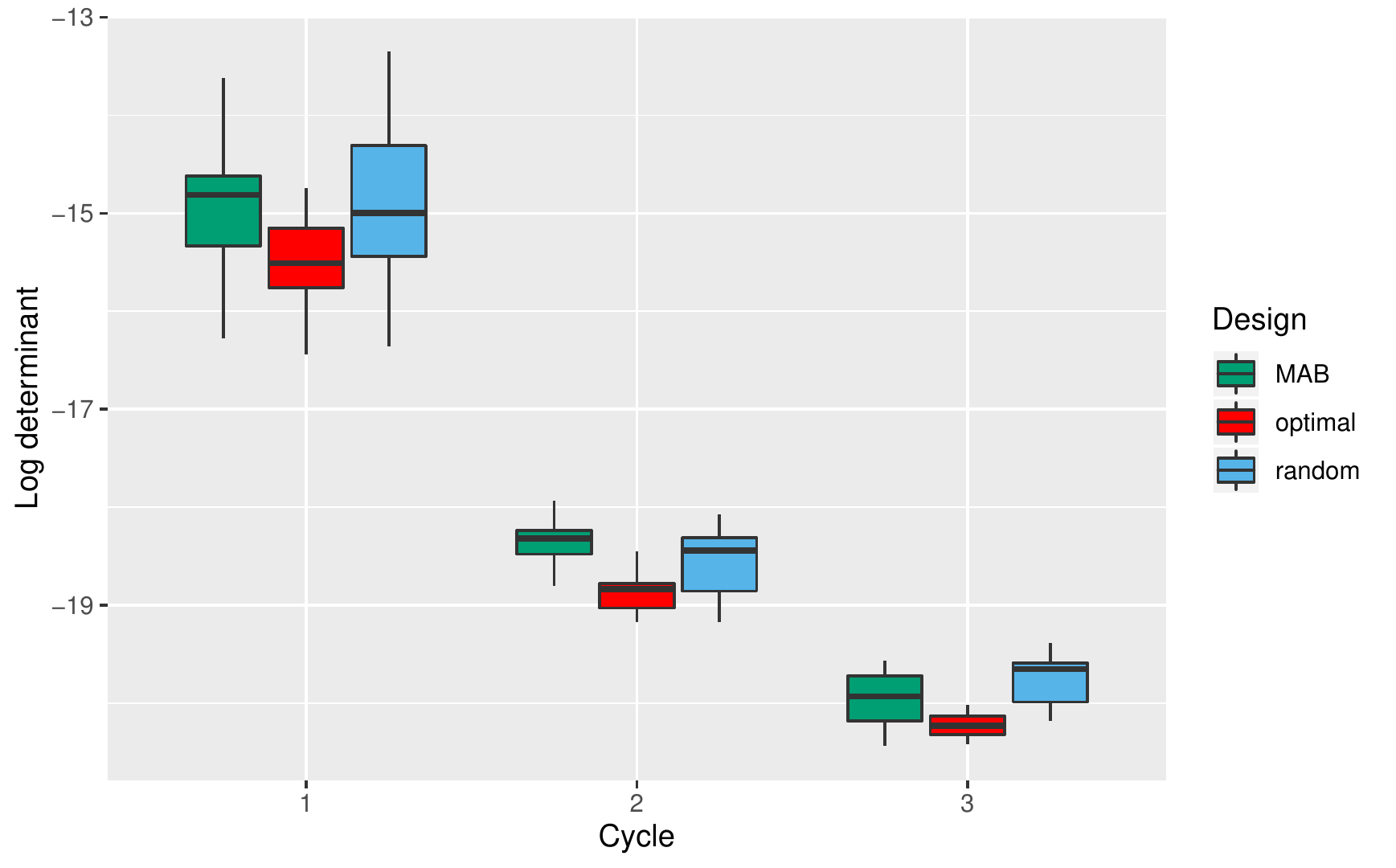} \vspace{-.3cm}
\captionsetup{type=figure}
\captionof{figure}{The boxplot of the distribution of the log-determinant of the posterior variance-covariance matrix for each design after each treatment cycle over 20 simulations from Example 2.}
\label{fig:Log_det_E2}
\end{center}

\begin{center}
\includegraphics[width=16cm]{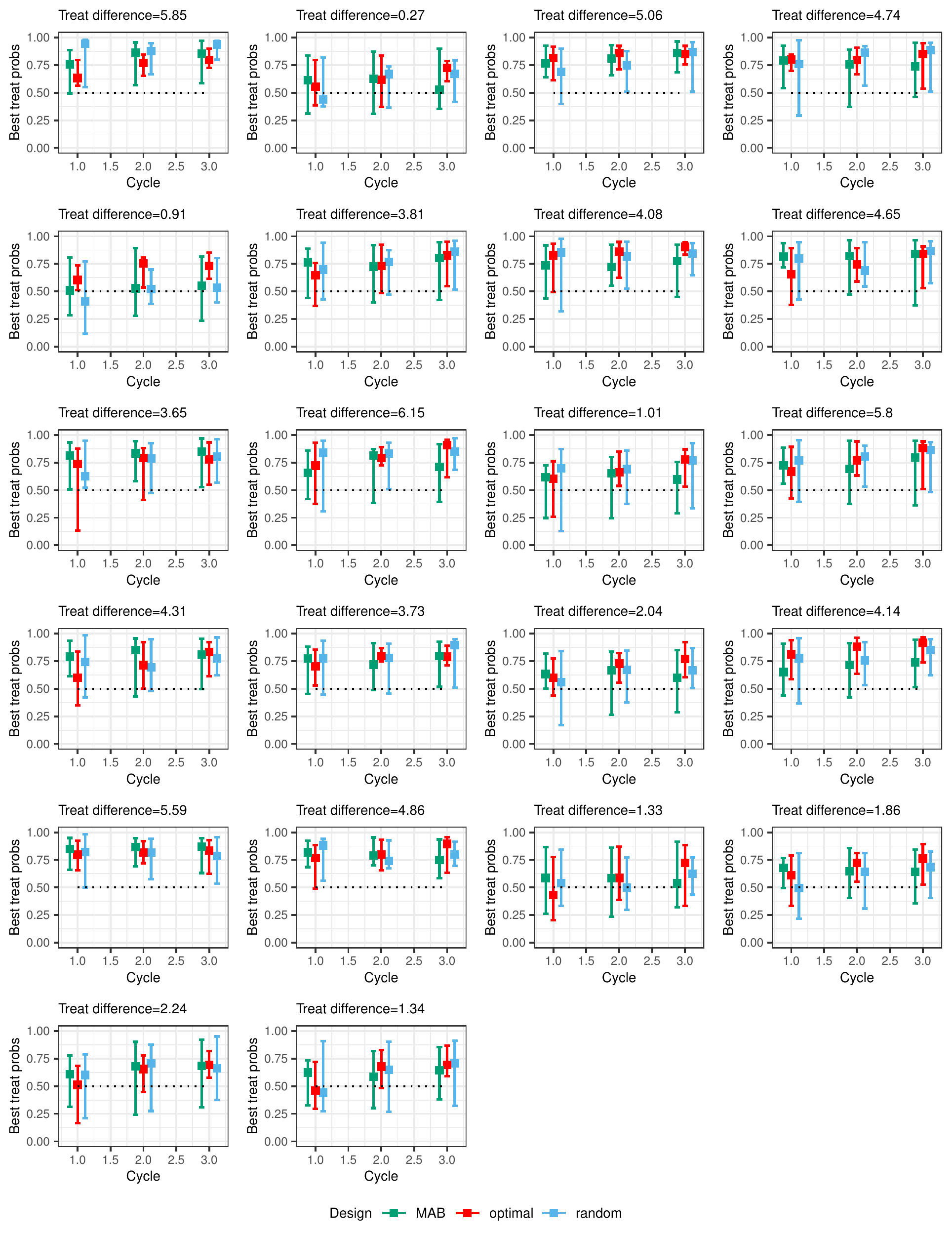} \vspace{-.2cm}
\captionsetup{type=figure}
\captionof{figure}{The probability (with $95\%$ credible intervals) of identifying the best treatment for each patient after each cycle when treatments were assigned using optimal, MAB and randomised N-of-1 designs from Example 2.}
\label{fig:Best_Treat_probsE2}
\end{center}

\begin{center}
\includegraphics[width=16cm]{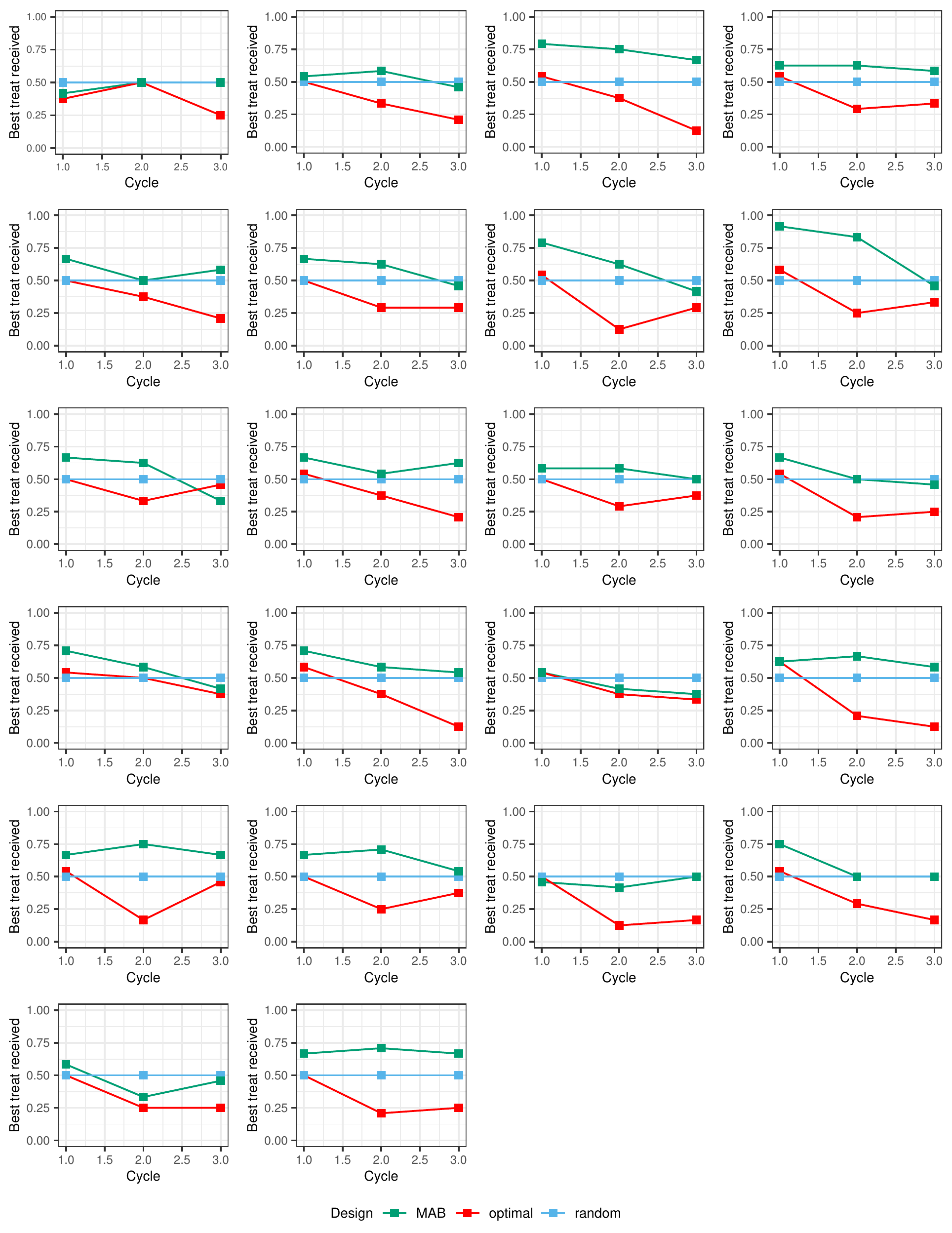} \vspace{-.2cm}
\captionsetup{type=figure}
\captionof{figure}{The proportion of time that their best treatment was received for each patient in each treatment cycle when treatments were assigned using optimal, MAB and randomised N-of-1 designs over 20 simulations for three cycles from Example 2.}
\label{fig:Best_Treat_receivedE2}
\end{center}

\section{Discussion}

In this work, we have developed a Bayesian adaptive design approach to find optimal treatment allocations for N-of-1 trials. As was seen, the designs derived from our approach can be used to determine the best treatment assignment for each patient with a fewer number of treatment cycles and/or provide more certainty about this after three cycles (typical duration of an N-of-1 trial). The empirical evidence presented in this paper demonstrates that the proposed Bayesian adaptive framework can be used to efficiently estimate both population parameters and the random effects in realistically sized N-of-1 trials, and benefits of this approach over alternatives were demonstrated. As such, we propose this method could be adopted in future N-of-1 trials to determine appropriateness in real-world settings.

In the first example, four design scenarios were considered to investigate the performance of our adaptive design approach under different parameter settings. It was found that the optimal designs were preferred for estimating the population parameters and the random effects when compared to the MAB and randomised N-of-1 designs. Furthermore, in using the optimal designs, we were able to determine the best treatment assignment for each patient in a fewer number of treatment cycles. When there was considerable variability in the random effects and a small difference between the individual treatment effects, it was difficult to determine the best treatment assignment with our optimal design approach but this was also observed for the other two approaches considered in this paper. In all four scenarios, MAB designs chose the best treatment for each patient a larger number of times than the optimal and randomised N-of-1 designs but we note that this did not translate into more certainty about which treatment was best for each patient.

When we considered the motivating example for this research, benefits were seen in adopting our optimal design approach compared to the MAB and randomised N-of-1 designs in terms of estimating model parameters. Of note, this again translated into more certainty about the preferred treatment for each patient during and at the end of the study.  Thus, it seems as though more information from the N-of-1 study of MPH could have been obtained if Bayesian adaptive design methods were implemented. This was also seen when an alternative distribution was considered for the response (see Section B of the supplementary material). Accordingly, we hope to explore the use of our methods in real N-of-1 trials into the future.

The two Laplace approximations proposed in this paper to form an approximation to the posterior distribution of the parameter is different to what has previously been proposed in the literature by, for example, \cite{Overstall2018}. In such work, authors have proposed a single Laplace approximation formed by considering the conditional likelihood i.e.\ not integrating out the random effects.  Given this, it is of interest to compare the two approaches to determine which appears to yield a better approximation to the posterior distribution.  To investigate this, a separate simulation study was undertaken where posterior distributions obtained from both approaches were compared to that obtained from MCMC. For this, the data generating model defined in Example 1 with five patients was used to simulate 50 data sets, each based on a typical N-of-1 design with 3 cycles. For each simulated data set, MCMC and the two Laplace-based approaches were used to form an approximation to the posterior distribution of the parameters. The posterior mean and variance for all parameters were recorded for each simulation, and the distribution of these is summarised in Table \ref{tab:Sim_study_new}. As can be seen, the posterior distributions obtained from MCMC and the Laplace approach proposed in this paper have similar mean values, while the approximation from \cite{Overstall2018} are noticeably different, particularly for the random effect parameters.  Both Laplace approximations appear to underestimate the variance of the parameters but this is much more apparent when the approach from \cite{Overstall2018} is used.  An example of the posterior distributions obtained under these three approaches is shown in Figure  \ref{fig:Posterior_comparison_NewEx}. The discussed advantages of our Laplace methods are highlighted in this plot.

As discussed throughout this paper, standard N-of-1 trials randomise treatment allocations within cycles. However, such randomisation has no place in a Bayesian decision-theoretic framework. In essence, a randomised decision cannot result in a higher expected utility than a deterministic decision. Notwithstanding this, it is recognised that there may be unobserved confounding variables and that randomisation protects against the effects of these (see \cite{soton46376}). Hence it may be desirable to randomly allocate treatments to patients. Within our implementation, approaches for randomising treatment allocation based on the expected amount of information to be gained could be implemented (e.g. \cite{Anthony2005}; \cite{Steffen2019}).

Future development of our adaptive design approach could include extensions to other types of trials.  Of note, cross-over, single case experimental designs and step-wedge designs can be viewed as special cases of the N-of-1 trial design.  Thus, our approach could potentially be adopted within such settings.  Further, it would be interesting to explore our methodologies for designing platform trials where patients can be allocated to different treatments over time \citep{Berry2015}. Such trials generally consider more treatments when compared to N-of-1s, and the availability of different treatments can varying depending on the patient.  Our approach to targeting information at the population and individual patient level could prove useful in, for example, quickly discounting ineffective treatments.  We plan to explore this in future endeavours. 

When designing N-of-1 trials, typically there is some information available about treatment effects from previously collected data.  Such prior information can be used to determine an appropriate model for which to undertaken adaptive design.  However, if not sufficient, then uncertainty at the model level should be incorporated into the selection of treatments.  Of note, the identified Bayesian adaptive design approach can be extended to incorporate model uncertainty by forming a non-trivial prior on the model space. Further, in cases where it is desirable to (additionally) learn which model is most appropriate for the data, dual-purpose utility functions could be considered \citep{borth_1975,mcgree2016developments}. In cases where different data types and multiple primary and secondary outcomes are considered, extensions to the work of \cite{Senarathne2020} may prove fruitful in forming a more accurate approximation to the posterior distribution. These are also areas of research which we hope to explore in the future.

\section*{Acknowledgement}

SGJS was supported by QUTPRA scholarship from the Queensland University of Technology. Computational resources and services used in this work were provided by the HPC and Research Support Group, Queensland University of Technology, Brisbane, Australia.

\bibliographystyle{agsm}
\small {
}

\section*{Appendix}
\subsection*{Comparison between our proposed Laplace approach, MCMC and a Laplace approach based on the conditional likelihood}

\renewcommand{\thefigure}{A\arabic{figure}}
\renewcommand{\thetable}{A\arabic{table}}
\setcounter{figure}{0} 
\setcounter{table}{0} 

\begin{center}
	\centering
\includegraphics[width=14cm]{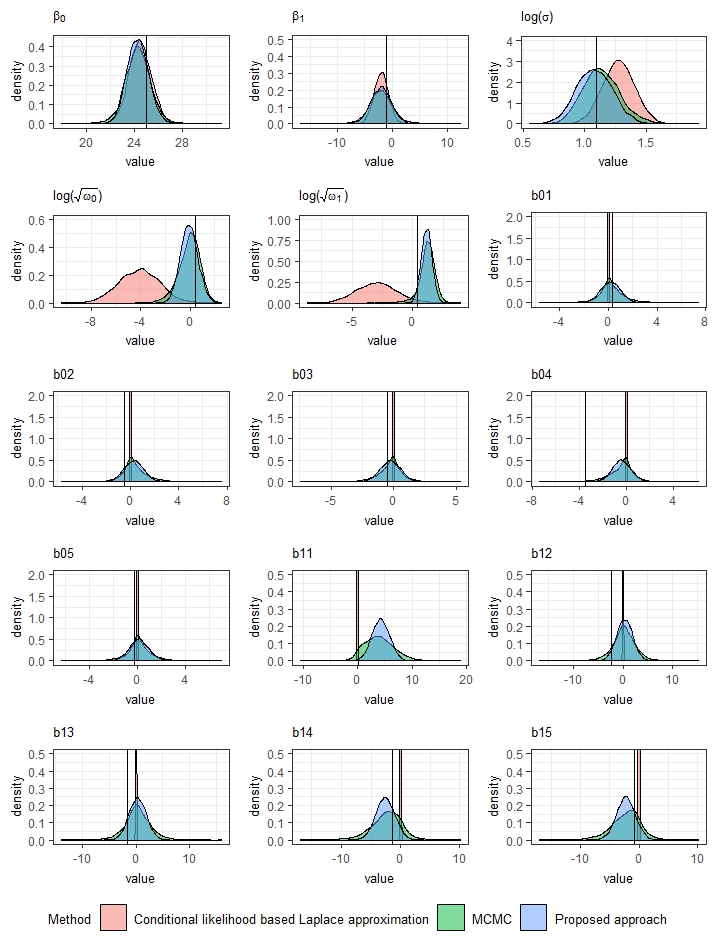} \vspace{0cm}
\captionsetup{type=figure}
\captionof{figure}{The posterior distributions obtained from our proposed Laplace approximation, MCMC, and a Laplace approach based on the conditional likelihood for Example 1 with 5 patients.}
\label{fig:Posterior_comparison_NewEx}
\end{center}

\begin{landscape}
\begin{table}[b]
\caption {A comparison between the posterior distributions obtained from our proposed approach, an MCMC approach and the conditional likelihood based Laplace approximation proposed in the literature} \label{tab:Sim_study_new} \vspace{.6cm}
\begin{tabular}{crrrrrr}
\hline
\multirow{2}{*}{Parameter} & \multicolumn{2}{c}{\begin{tabular}[c]{@{}c@{}}Likelihood based Laplace approximation \\ (proposed approach)\end{tabular}} & \multicolumn{2}{c}{MCMC approach} & \multicolumn{2}{c}{\begin{tabular}[c]{@{}c@{}}Conditional likelihood based Laplace \\ approximation\end{tabular}} \\ \cline{2-7} 
 & \multicolumn{1}{l}{\begin{tabular}[c]{@{}l@{}}Posterior mean \\ with 95\% \\ credible intervals\end{tabular}} & \multicolumn{1}{l}{\begin{tabular}[c]{@{}l@{}}Posterior variance\\ with 95\% \\ credible intervals\end{tabular}} & \multicolumn{1}{l}{\begin{tabular}[c]{@{}l@{}}Posterior mean \\ with 95\% \\ credible intervals\end{tabular}} & \multicolumn{1}{l}{\begin{tabular}[c]{@{}l@{}}Posterior variance\\ with 95\% \\ credible intervals\end{tabular}} & \multicolumn{1}{l}{\begin{tabular}[c]{@{}l@{}}Posterior mean \\ with 95\% \\ credible intervals\end{tabular}} & \multicolumn{1}{l}{\begin{tabular}[c]{@{}l@{}}Posterior variance \\ with 95\% \\ credible intervals\end{tabular}} \\ \hline
$\beta_0$ & 23.98 (23.00, 25.03) & 1.16 (0.64, 1.92) & 23.97 (22.99, 25.03) & 2.25 (1.04, 4.47) & 23.98 (22.94, 25.13) & 0.88 (0.37, 1.47) \\ \hline
$\beta_1$ & -1.94 (-3.4, -0.29) & 1.93 (1.14, 3.35) & -1.93 (-3.42, -0.29) & 3.44 (1.81, 6.64) & -1.94 (-3.34, -0.32) & 1.45 (0.73, 2.69) \\ \hline
$\log(\sigma)$ & 1.04 (0.75, 1.25) & 0.02 (0.02, 0.03) & 1.09 (0.80, 1.30) & 0.02 (0.02, 0.03) & 1.07 (0.71, 1.40) & 0.02 (0.02, 0.02) \\ \hline
$\log{(\sqrt{\omega_0})}$ & 0.42 (-0.24, 0.98) & 0.32 (0.17, 0.49) & 0.46 (-0.22, 1.16) & 0.57 (0.29, 0.74) & -2.59 (-4.16, 0.82) & 1.86 (0.17, 2.67) \\ \hline
$\log{(\sqrt{\omega_1})}$ & 0.58 (0.14, 1.18) & 0.41 (0.17, 0.54) & 0.56 (0.10, 1.34) & 0.67 (0.33, 0.81) & -2.31 (-2.99, 1.05) & 2.35 (0.14, 2.75) \\ \hline
b01 & 1.17 (-0.03, 3.17) & 1.01 (0.46, 1.67) & 1.15 (-0.01, 3.11) & 2.84 (1.04, 5.63) & 0.77 (0.00, 3.2) & 0.59 (0.00, 2.01) \\ \hline
b02 & 0.04 (-1.12, 1.19) & 1.00 (0.46, 1.66) & 0.05 (-1.08, 1.18) & 2.65 (0.98, 5.18) & -0.07 (-1.63, 0.81) & 0.52 (0.00, 1.80) \\ \hline
b03 & 0.20 (-1.15, 1.18) & 1.00 (0.46, 1.66) & 0.20 (-1.21, 1.16) & 2.62 (0.99, 5.24) & 0.10 (-0.59, 1.20) & 0.50 (0.00, 1.86) \\ \hline
b04 & -1.65 (-3.65, -0.21) & 1.01 (0.47, 1.70) & -1.62 (-3.67, -0.24) & 3.06 (0.98, 5.64) & -0.91 (-3.39, 0.00) & 0.64 (0.00, 2.23) \\ \hline
b05 & 0.25 (-0.72, 1.34) & 1.00 (0.47, 1.66) & 0.25 (-0.67, 1.27) & 2.61 (1.00, 5.10) & 0.12 (-0.19, 0.88) & 0.50 (0.00, 1.81) \\ \hline
b11 & 1.19 (-0.51, 3.96) & 1.73 (0.96, 3.03) & 1.18 (-0.59, 3.89) & 4.38 (1.93, 8.71) & 0.53 (0.00, 3.96) & 0.44 (0.00, 3.64) \\ \hline
b12 & -0.82 (-3.21, 0.57) & 1.73 (0.96, 2.93) & -0.82 (-3.20, 0.61) & 4.18 (1.90, 8.34) & -0.33 (-3.01, 0.00) & 0.40 (0.00, 3.00) \\ \hline
b13 & 0.04 (-1.78, 1.57) & 1.74 (0.95, 3.10) & 0.04 (-1.81, 1.58) & 4.02 (1.84, 7.69) & 0.01 (-0.49, 0.38) & 0.42 (0.00, 3.09) \\ \hline
b14 & -0.81 (-3.56, 0.73) & 1.72 (0.98, 3.03) & -0.82 (-3.42, 0.82) & 4.39 (1.87, 9.35) & -0.34 (-3.20, 0.00) & 0.43 (0.00, 3.13) \\ \hline
b15 & 0.39 (-1.19, 2.07) & 1.75 (0.98, 3.06) & 0.38 (-1.27, 1.90) & 4.00 (1.85, 8.00) & 0.18 (0.00, 1.28) & 0.40 (0.00, 3.01) \\ \hline
\end{tabular}
\end{table}
\end{landscape}

\includepdf[pages=-]{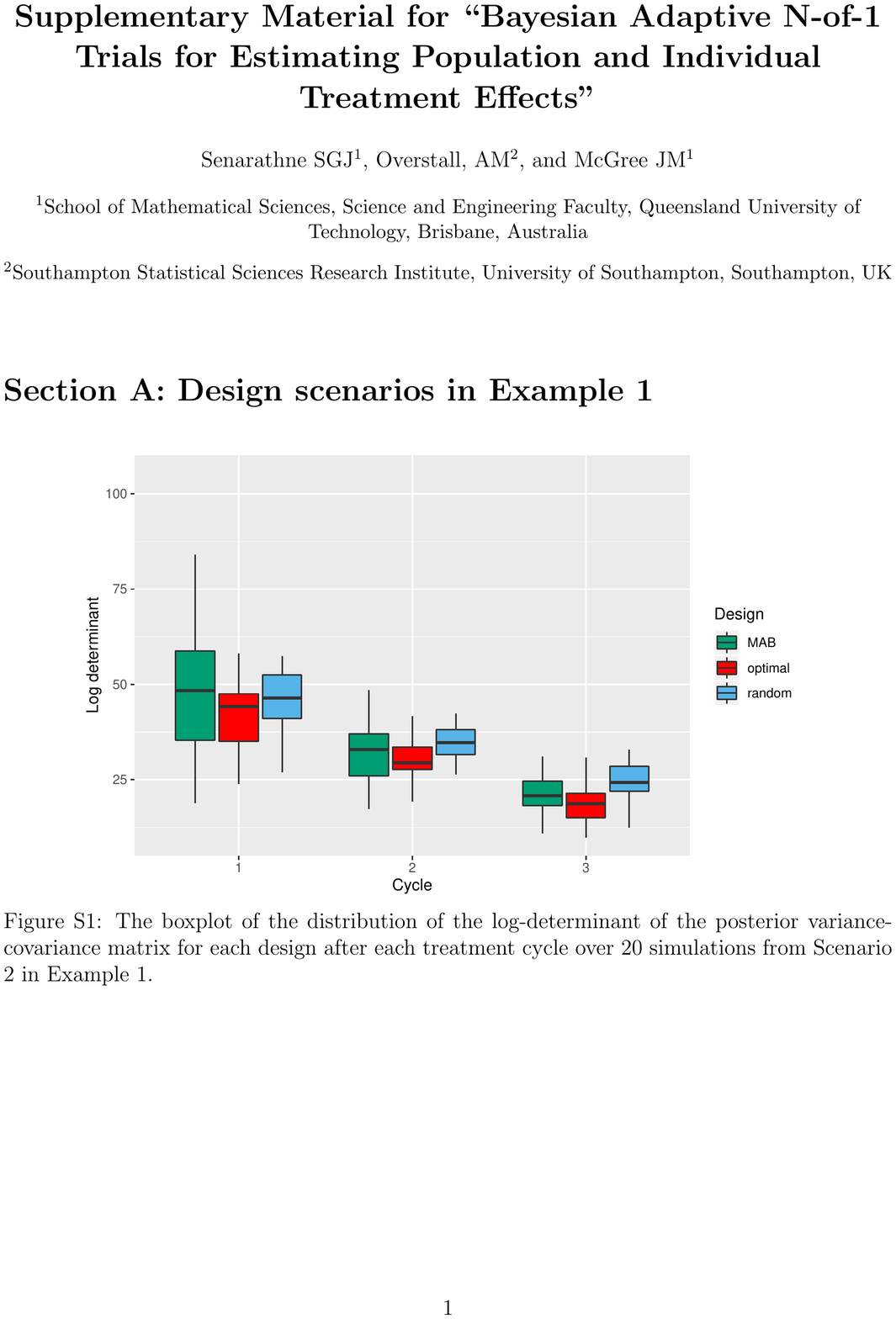}

\end{document}